\documentstyle[12pt,psfig]{article}
\newcommand{\mysection}{\setcounter{equation}{0}\section}

\def\beq{\begin{equation}}
\def\eeq{\end{equation}}
\def\beqa{\begin{eqnarray}}
\def\eeqa{\end{eqnarray}}

\begin{document}
\begin {flushright}
FSU-HEP-991216\\
\end {flushright} 
\vspace{3mm}
\begin{center}
{\Large \bf Soft-gluon resummation and NNLO corrections
for direct photon production}
\end{center}
\vspace{2mm}
\begin{center}
{\large Nikolaos Kidonakis and J.F. Owens}\\
\vspace{2mm}
{\it Physics Department\\
Florida State University\\
Tallahassee, FL 32306-4350, USA} \\
\end{center}

\begin{abstract}
The resummation of threshold logarithms for 
direct photon production cross sections in hadronic collisions is presented. 
The resummation is based on the factorization properties of the cross
section and is formulated at next-to-leading logarithmic or higher
accuracy. Full analytical and numerical results for 
the next-to-next-to-leading order expansion of the resummed cross section are 
given. A substantial reduction of factorization scale dependence is observed.
A comparison to experimental results from the E-706 and UA-6 
experiments is presented.

\end{abstract}
\pagebreak

\mysection{Introduction}

The production of photons with large transverse momenta has long been 
recognized as an important probe of hard-scattering dynamics. In particular, 
the subprocess $q g \rightarrow \gamma q$ offers sensitivity to the gluon  
distribution function in lowest order. Thus, direct photon production is 
often cited as having the potential to place strong constraints on the gluon 
distribution in global fits of parton distributions. In recent years, 
however, the situation has become more complex. Recent detailed 
phenomenological studies have shown that it is not possible to describe the 
existing set of direct photon experimental results using a next-to-leading 
order 
QCD calculation \cite{aurenche,dg1}. Even with the full freedom of 
separately adjusting the 
renormalization and the factorization scales for the distribution and 
fragmentation functions, one can not correctly describe all of the 
experimental results \cite{VOG}. In \cite {dg1} it was pointed out that for 
many of the data sets the 
theoretical predictions fall below the data near the low-$p_T$ end of the 
$p_T$ range spanned by the data. This has been interpreted \cite{kT} as 
providing evidence for the inclusion of $k_T$-smearing which is due to initial 
state radiation. However, this interpretation is not completely supported by 
all of the available data and the analysis of \cite{aurenche} shows that 
agreement with a subset of the data can be obtained if one considers a 
restricted set of data taken only on proton targets (so as to remove possible 
nuclear effects) and also if minimum $p_T$ cuts on the data sets are applied 
which restrict the comparisons to regions where the scale dependence of the 
theory is moderate, and, hence, the theoretical predictions are thought to be 
reliable. 

In light of the situation described in the preceeding paragraph, efforts are 
underway to improve the precision of the theoretical predictions for direct 
photon production. One such type of improvement involves the resummation of 
large logarithms which result from phase space limitations on real gluon 
emission near the threshold region for the underlying parton scattering 
subprocesses. Such corrections are expected to be important in the region of 
large photon transverse momentum, where $x_T=2p_T/{\sqrt S}$ approaches one. 
Nevertheless, it is of interest to examine the effects of such resummation 
calculations over an extended range of $x_T$. 

Recently, threshold resummation for direct photon production has
been discussed in Refs. \cite{LOS,CMN,CMNOV,DIS99}. 
In Refs. \cite{CMN,CMNOV}, a resummation 
calculation was presented for the inclusive $E_T$ distribution, 
$d \sigma/d E_T$, and compared with some of the available fixed target 
data. The results showed that, as expected, the corrections are large as 
$x_T$ approaches the edge of phase space. In addition, the scale dependence of 
the resulting distributions is reduced compared to that of the next-to-leading
order calculation. Finally, in the region containing the data the resummed 
cross section was nearly the same as the next-to-leading order cross section 
if the factorization and renormalization scales were chosen to be 
$\mu = p_T/2.$

Another formalism exists for treating the resummation of threshold logarithms 
in inclusive processes, a review of which can be found in Ref. \cite{NK}.
This formalism has been applied to lepton pair, heavy quark, and dijet 
production at fixed invariant mass and more recently has
been extended to single-particle inclusive cross sections \cite{LOS},
including direct photon \cite{LOS} and $W$ + jet production \cite{NKVD,HEP99}.
In the present paper this formalism is applied to the inclusive 
invariant cross section for direct photon production. One important
feature of this method is that one directly obtains the angular dependence 
(or rapidity dependence) for the produced photon. The resummation explicitly 
includes all the 
leading and next-to-leading logarithms (NLL). For detailed numerical work and 
comparison to other calculations, the resummed cross section expression 
is expanded to next-to-leading order (NLO) and next-to-next-to-leading order 
(NNLO). Matching the resulting expressions to existing complete NLO 
calculations yields the leading, next-to-leading, and next-to-next-to-leading 
logarithms (NNLL) through ${\cal O}(\alpha \alpha_s^3)$.
The NLO expansion agrees fully near partonic threshold with existing
exact NLO calculations \cite{dgamma1loop,GV} while our NNLO results 
provide new predictions.
A comparison with recent high statistics fixed target results is also 
presented. 

\mysection{Factorization and resummed cross section}

\subsection{Factorized cross section}

In this section we discuss the application of the threshold resummation 
formalism described in Ref. \cite{NK} to direct photon production. 
The self-contained presentation follows the techniques of 
Refs. \cite{LOS,NKVD}.

\begin{figure}
\centerline{
\psfig{file=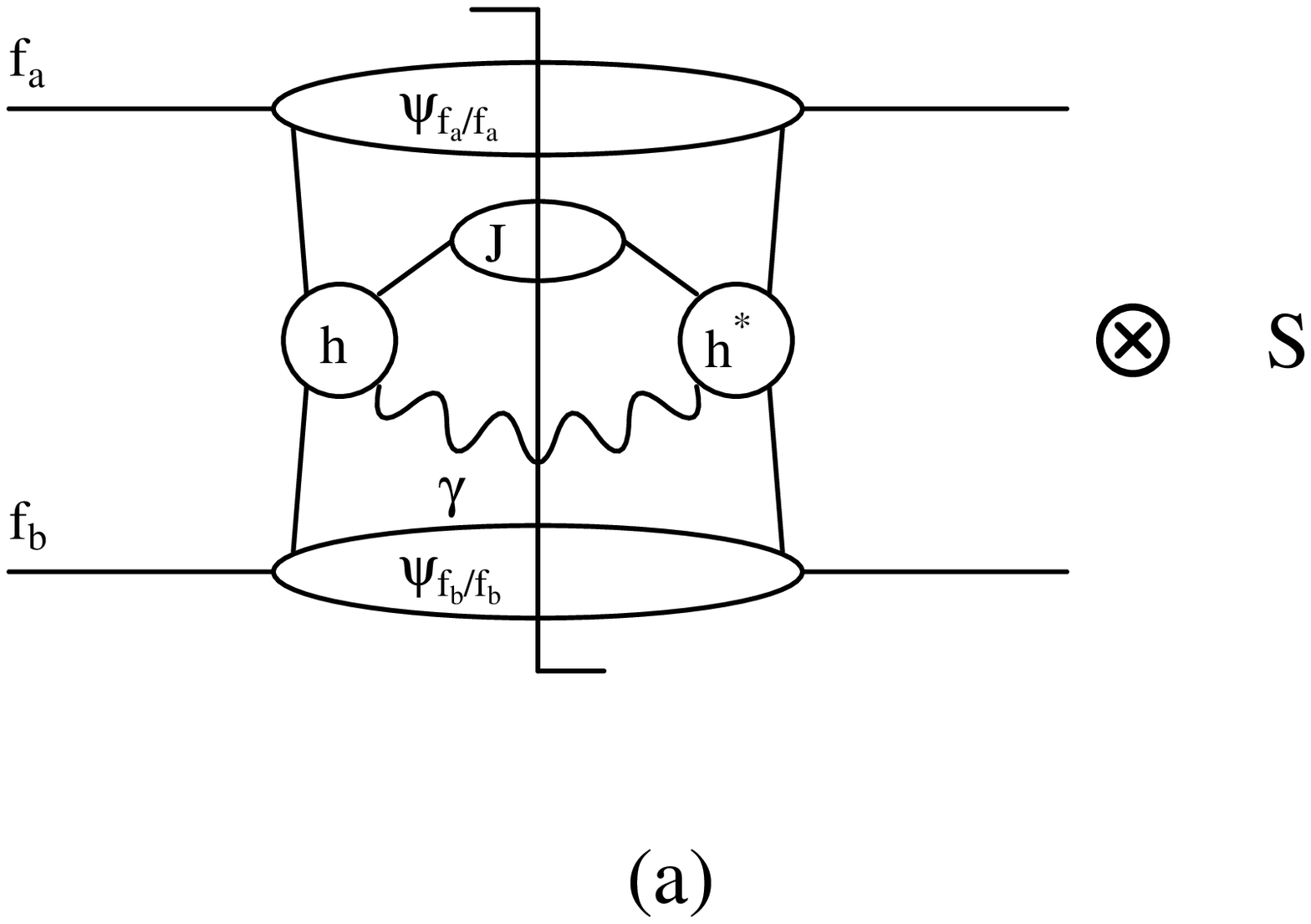,height=1.7in,width=2.5in,clip=}
\hspace{8mm}
\psfig{file=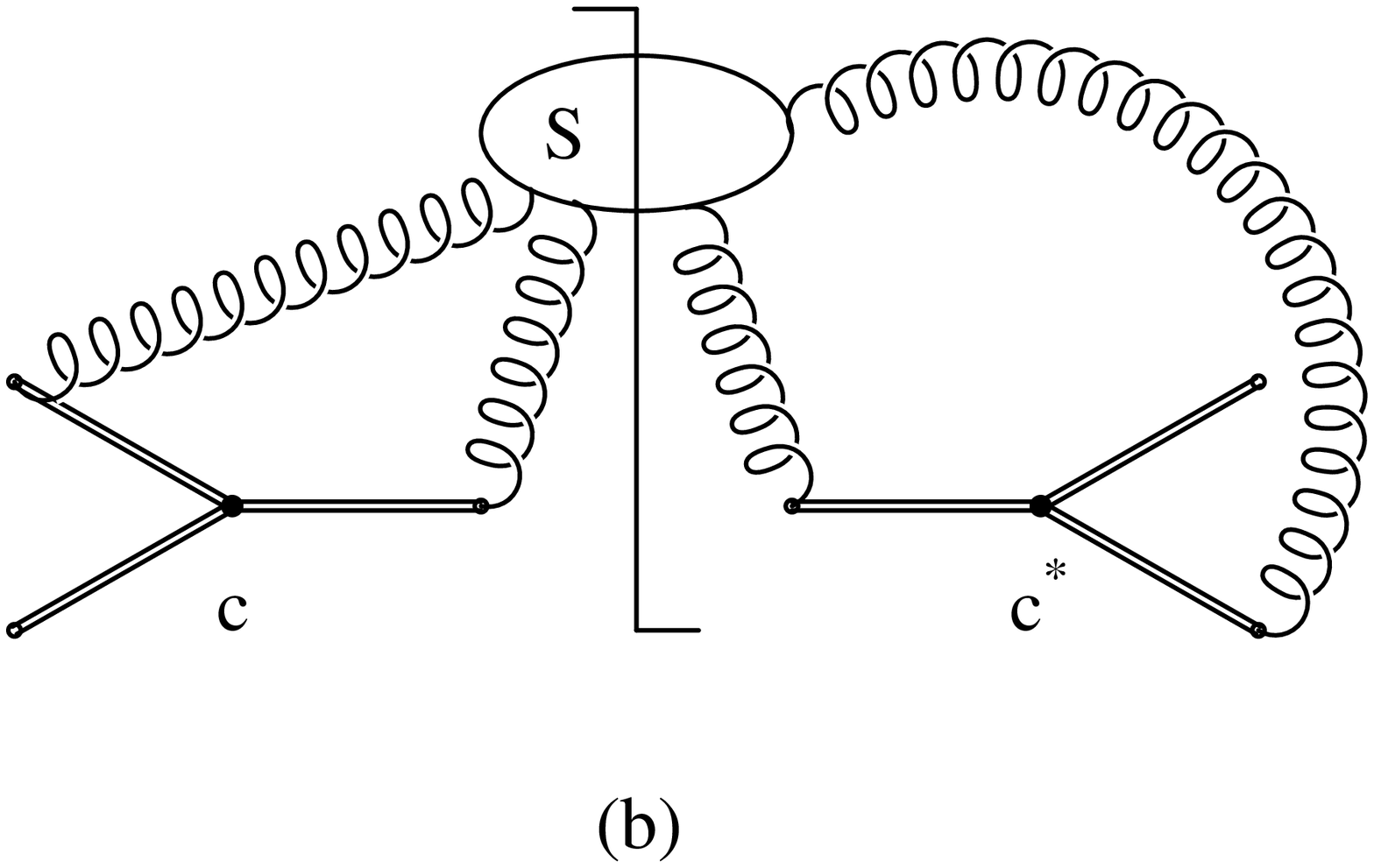,height=1.7in,width=2.5in,clip=}}
{Fig. 1. (a) Factorization for direct photon production 
near partonic threshold.
(b) The soft-gluon function $S$, in which
the vertices $c$ link ordered exponentials, representing the partons
in the hard scattering.}
\label{fig1}
\end{figure}

The quantity to be calculated is the inclusive invariant cross section for 
the production of photons in hadronic collisions:
\beq
h_A(p_A)+h_B(p_B) \rightarrow \gamma(p_{\gamma}) +X \, .
\eeq
The Mandelstam invariants formed form the hadron and photon four-vectors are
\beq
S=(p_A+p_B)^2 \, , \quad T=(p_A-p_{\gamma})^2 \, , \quad
U=(p_B-p_{\gamma})^2 \, .
\eeq
The parton-parton scattering  subprocesses contributing to direct photon 
production in lowest order are
\beq
q(p_a)+g(p_b) \longrightarrow  \gamma(p_{\gamma}) + q(p_J)
\eeq
and
\beq
q(p_a)+{\bar q}(p_b) \longrightarrow \gamma(p_{\gamma}) + g(p_J)\, .
\eeq
The corresponding Mandelstam invariants constructed from the parton and 
photon four-vectors are given by
\beq
s=(p_a+p_b)^2 \, , \quad t=(p_a-p_{\gamma})^2 \, , 
\quad u=(p_b-p_{\gamma})^2 \, ,
\eeq
which satisfy $s_4 \equiv s+t+u=0$ at threshold.
The variable $s_4$ is the square of the invariant mass of the system
recoiling against the observed photon. It will also be useful to 
define a corresponding variable, $S_4=S+T+U$, using the hadronic Mandelstam 
variables.

The factorized form of the cross section for direct photon 
production is a convolution of the parton distribution functions
$\phi$ with the parton-parton hard scattering cross section:
\beqa
E_{\gamma}\frac{d^3\sigma_{h_Ah_B\rightarrow\gamma X}}{d^3 p_{\gamma}}
&=&\sum_{f} \int dx_a dx_b \, \phi_{f_a/h_A}(x_a,\mu_F^2) \,
\phi_{f_b/h_B}(x_b,\mu_F^2)
\nonumber \\ &&  \times \, 
E_{\gamma} \frac{d^3{\hat \sigma}_{f_a f_b \rightarrow \gamma X}}
{d^3  p_{\gamma}}(s,t,u,\mu_F,\alpha_s(\mu_R^2)) \, .
\label{factdp}
\eeqa
The initial-state
collinear singularities have been factorized and absorbed into the
$\phi$'s at a factorization scale $\mu_F$, while $\mu_R$ is the
renormalization scale.

The threshold for the parton-parton scattering subprocesses occurs at 
$s_4=0$. At this point the photon recoils against a massless parton and 
receives the maximum amount (one half) of the parton center of mass energy 
allowed. The values of $x_a$ and $x_b$ corresponding to this point are the 
minimum values which can give rise to a photon of the specified rapidity and 
transverse momentum. Near threshold the phase space for the emission of 
real gluons is limited and there is an incomplete cancellation of
infrared divergences against 
virtual gluon emission contributions. Although the remainder is finite, 
there are large logarithmic corrections. In general, $\hat{\sigma}$ 
includes plus distributions with respect
to $s_4$ at $n$th order in $\alpha_s$ of the type
\beq
\left[\frac{\ln^{m}(s_4/p_T^2)}{s_4} \right]_+, \hspace{10mm} m\le 2n-1\, ,
\label{s4plus}
\eeq
where $p_T=(tu/s)^{1/2}$ is the transverse momentum of the photon.

To organize these plus distributions we introduce a refactorization 
in terms of new parton distributions $\psi$ and a jet function $J$ that 
absorb the collinear singularities from the incoming partons and outgoing
jet \cite{LOS,KS,KOS}, respectively; a soft gluon function $S$ that
describes noncollinear soft gluon emission \cite{KS,KOS}; and a
short-distance hard scattering function $H$.
This refactorization is shown in Fig. 1 for the partonic cross section 
with the colliding hadrons in Eq. (\ref{factdp}) replaced by partons  
$f_a$, $f_b$. In our soft-gluon approximation
the partons in the hard scattering are represented by eikonal lines
(ordered exponentials). The cut diagram in Fig. 1 shows
contributions from the amplitude and its complex conjugate,
with $H=h h^*$.

The momenta of the parton lines entering the hard scattering
subprocess are then $x_a p_A$ and $x_b p_B$, where now $p_A$ and 
$p_B$ denote the momenta of external partons.
If soft radiation of total momentum $k_S$ is emitted, 
then momentum conservation gives
$x_a p_A+x_b p_B=p_{\gamma}+p_J+k_S$ where $p_J$ represents the momentum 
of everything in the final state excluding the photon and the soft radiation. 
Using the definitions of the Mandelstam variables presented previously, 
and neglecting corrections of order $S_4^2$, the relation between $S_4$ and 
$s_4$ can be expressed as \cite{LOS}
\beqa
\frac{S_4}{S}&=&\frac{1}{S}\left[2 p_J \cdot k_S + p_J^2 
+(1-x_a) 2p_A \cdot p_J+(1-x_b) 2p_B \cdot p_J\right]
\nonumber \\
&\equiv&w_s+w_J+w_a\frac{(-u)}{s}+w_b\frac{(-t)}{s} 
\nonumber \\
&=&\frac{s_4}{s}-(1-x_a)\frac{u}{s}
-(1-x_b)\frac{t}{s} \, ,
\label{weights}
\eeqa 
where the dimensionless $w_i$'s measure the fractional contributions 
to $S_4/S$ of the gluon emission associated with 
the functions $\psi_i$, $J$, and $S$. 
The refactorized partonic cross section can be written as
a convolution \cite{LOS}
\beqa
E_{\gamma}\frac{d^3\sigma_{f_af_b\rightarrow\gamma X}}{d^3 p_{\gamma}}&=&
H \int dw_a \, dw_b \, dw_J \, dw_S \,  
\psi_{f_a/f_a}(w_a) \, \psi_{f_b/f_b}(w_b) \, J(w_J)
\nonumber \\ && \hspace{-20mm} \times \, 
S(w_S {\sqrt s}/{\mu_F}) \; 
\delta\left(\frac{S_4}{S}-w_s-w_J
-w_a\frac{(-u)}{s}-w_b\frac{(-t)}{s}\right) \, .
\eeqa
If we take moments of the above equation, with $N$ the moment
variable, we can write the partonic cross section as
\beqa
\int \frac{dS_4}{S} {\rm e}^{-NS_4/S}
E_{\gamma}\frac{d^3{\sigma}_{f_a f_b\rightarrow \gamma X}}
{d^3 p_{\gamma}}
&=&{\tilde{\psi}}_{f_a/f_a}(N_a)\, {\tilde{\psi}}_{f_b/f_b}(N_b)\,
{\tilde{J}}(N) \, H \, {\tilde{S}}({\sqrt s}/(N\mu_F)) \, ,
\nonumber \\
\label{sigmamomdp}
\eeqa
where 
\beq
{\tilde{\psi}}_{f_a/f_a}(N_a)=
\int dw_a {\rm e}^{-N_a w_a} {\psi}_{f_a/f_a}(w_a)\, ,
\eeq
and similarly for the other functions, 
with $N_a=N(-u/s)$ and $N_b=N(-t/s)$ for either
partonic subprocess.

Next, we take moments of Eq. (\ref{factdp}) with the incoming hadrons
replaced by partons and, using the last line of Eq. (\ref{weights}), we find 
\beqa
\int \frac{dS_4}{S} {\rm e}^{-NS_4/S}
E_{\gamma}\frac{d^3\sigma_{f_af_b\rightarrow\gamma X}}{d^3 p_{\gamma}}&=&
\int dx_a \, {\rm e}^{-N(-u/s)(1-x_a)} \phi_{f_a/f_a}(x_a,\mu_F^2) \,
\nonumber \\ && \hspace{-47mm} \times
\int dx_b \, {\rm e}^{-N(-t/s)(1-x_b)} \phi_{f_b/f_b}(x_b,\mu_F^2) \,
\int \frac{ds_4}{s} \, {\rm e}^{-Ns_4/s} E_{\gamma}
\frac{d^3{\hat \sigma}_{f_af_b\rightarrow\gamma X}(s_4)}{d^3 p_{\gamma}}
\nonumber \\ && \hspace{-15mm}
\equiv {\tilde \phi}_{f_a/f_a}(N_a) {\tilde \phi}_{f_b/f_b}(N_b) E_{\gamma} 
\frac{d^3{\hat \sigma}_{f_af_b\rightarrow\gamma X}(N)}{d^3 p_{\gamma}} \, .
\label{moms4}
\eeqa
Comparing Eqs. (\ref{moms4}) and (\ref{sigmamomdp}) we may then
solve for the moments of the perturbative cross section $\hat \sigma$:
\beq
E_{\gamma}\frac{d^3{\hat \sigma}_{f_a f_b \rightarrow \gamma X}(N)}
{d^3 p_{\gamma}}
=\frac{{\tilde{\psi}}_{f_a/f_a}(N_a)\, {\tilde{\psi}}_{f_b/f_b}(N_b)}
{{\tilde{\phi}}_{f_a/f_a}(N_a)\,  {\tilde{\phi}}_{f_b/f_b}(N_b)}
{\tilde{J}}(N) \, H \, {\tilde{S}}({\sqrt s}/(N\mu_F)) \, .
\label{sigfact}
\eeq
The moments of the plus distributions with respect to $s_4$ 
in $\hat \sigma$ give powers of
$\ln N$ as high as $\ln^{2n}N$. In the next section these logarithms are 
resummed to all
orders in perturbation theory.

\subsection{Resummed cross section}

The resummation of the $N$-dependence of each of the functions in the
refactorized cross section, Eq. (\ref{sigfact}), 
is based on their renormalization properties
and has been extensively discussed in Refs. \cite{LOS,NKVD,KS,KOS,CLS}.
The resummed cross section in the $\overline{\rm MS}$ factorization 
scheme is given by:
\beqa
E_{\gamma}\frac{d^3{\hat{\sigma}}_{f_a f_b \rightarrow \gamma X}(N)}
{d^3 p_{\gamma}} &=&  
\exp \left \{ \sum_{i=a,b} \left [E^{(f_i)}(N_i)\right.\right. 
\nonumber\\ &&  \left. \left. 
{}-2\int_{\mu_F}^{2 p_i \cdot \zeta}{d\mu'\over\mu'}\; 
\left [\gamma_{f_i}(\alpha_s(\mu'{}^2))-\gamma_{f_if_i}
(N_i,\alpha_s(\mu'{}^2)) 
\right] \right] \right\}
\nonumber \\ && \hspace{-10mm} \times \; 
\exp \left \{E'_{(J)}(N) \right\} \;
H\left(s,t,u,\alpha_s(\mu_F^2)\right) 
\nonumber \\ && \hspace{-10mm} \times \;   
{\tilde S}\left(\alpha_s(s/N^2)\right) \;
\exp \left[\int_{\mu_F}^{\sqrt{s}/N} {d\mu' \over \mu'} \, 
2 \, {\rm Re} \Gamma_S\left(\alpha_s(\mu'^2)\right)\right] \, ,
\nonumber \\
\label{rescrosec}
\eeqa 
where $\zeta^{\mu}\equiv p_J^{\mu}/{\sqrt s}$.
The first exponent $E^{(f_i)}(N_i)$, which resums the $N$-dependence
of the ratio ${\tilde \psi}/{\tilde \phi}$, is given by \cite{KS,GS,CT1}
\beqa
E^{(f_i)}(N_i)
&=&
-\int^1_0 dz \frac{z^{N_i-1}-1}{1-z}\;
\left \{\int^1_{(1-z)^2} \frac{d\lambda}{\lambda}
A^{(f_i)}\left[\alpha_s\left(\lambda (2 p_i \cdot \zeta)^2\right)\right]\right.
\nonumber\\ &&  \hspace{10mm} \left.
{}+\frac{1}{2}\nu^{(f_i)}\left[\alpha_s((1-z)^2(2 p_i \cdot \zeta)^2)\right]
\right\} \, ,
\label{Eexp}
\eeqa
where, at next-to-leading order accuracy in $\ln N$,
we have 
\beq
A^{(f)}(\alpha_s) = C_f\left ( {\alpha_s\over \pi}
+\frac{1}{2} K \left({\alpha_s\over \pi}\right)^2\right )\, ,
\label{Aexp}
\eeq
and
\beq
\nu^{(f)}=2C_f\; {\alpha_s\over\pi}\, .
\label{nuf}
\eeq
Here $C_f=C_F=(N_c^2-1)/(2N_c)$ for an incoming quark,
and $C_f=C_A=N_c$ for an incoming gluon, with $N_c$ the number of colors,
while
\beq
K= C_A\; \left ( {67\over 18}-{\pi^2\over 6 }\right ) - {5\over 9}n_f\, ,
\eeq
where $n_f$ is the number of quark flavors \cite{KoTr}.

The anomalous dimensions $\gamma_f$ and $\gamma_{ff}$ of 
the fields $\psi$ and $\phi$ \cite{NKVD,KOS,LM} are given by
\beqa
\gamma_q &=& {3\over 4} C_F {\alpha_s\over\pi}\,; \qquad
\gamma_{qq} = - \left(\ln{N} - {3\over 4}\right) C_F {\alpha_s\over\pi} \, ,
\nonumber \\ \gamma_g &=& {\beta_0\over 4} {\alpha_s\over\pi}\,; \qquad
\gamma_{gg} = - \left(C_A \ln{N}-{\beta_0\over 4}\right){\alpha_s\over\pi}\, ,
\eeqa
for quark and gluon jets, respectively, where
$\beta_0= (11C_A-2n_f)/3$ is the one-loop coefficient of the $\beta$
function.

The exponent $E'_{(J)}$, which resums the $N$-dependence
of the final-state jet, is given by \cite{LOS,NKVD,KOS}
\beqa
E'_{(J)}(N)
&=&
\int^1_0 dz \frac{z^{N-1}-1}{1-z}\;
\left \{\int^{(1-z)}_{(1-z)^2} \frac{d\lambda}{\lambda}
A_{(J)}\left[\alpha_s(\lambda s)\right] \right.
\nonumber\\ &&  \left.
{}+B'_{(J)}\left[\alpha_s((1-z) s) \right] 
+B''_{(J)}\left[\alpha_s((1-z)^2 s) \right]\right\}\, ,
\label{Eprexp}
\eeqa
with $A_{(J)}$ given by Eq. (\ref{Aexp}) and $B'_{(J)}$, $B''_{(J)}$
given for quarks by~\cite{LOS,NKVD}
\beq
B'_{(q)}=\frac{\alpha_s}{\pi} \left(-\frac{3}{4}\right) C_F \, ,
\quad
B''_{(q)}=\frac{\alpha_s}{\pi} C_F \left[\ln(2\nu_q)-1\right] \, ,
\eeq
and for gluons by
\beq
B'_{(g)}=\frac{\alpha_s}{\pi} \left(- {\beta_0\over 4}\right) \, ,
\quad
B''_{(g)}=\frac{\alpha_s}{\pi} C_A \left[\ln(2\nu_g)-1\right] \, .
\eeq
The $\nu_i$ terms are gauge dependent. They are defined by
\beq
\nu_i \equiv \frac{(\beta_i \cdot n)^2}{|n|^2} \, ,
\eeq
where $\beta_i=p_i {\sqrt {2/s}}$ are the particle velocities 
and $n$ is the axial gauge vector, chosen so that $p_i\cdot \zeta=p_i \cdot n$
for $i=a,b$ \cite{LOS,NKVD}. 

\subsection{Soft anomalous dimensions}

The evolution of the soft function in Eq. (\ref{rescrosec}), which follows
from the renormalization group equation
\beq
\left(\mu\frac{\partial}{\partial\mu}+\beta(g)\frac{\partial}{{\partial}g}
\right)S=-2 ({\rm Re}\, \Gamma_S) \, S\, ,
\eeq
is given in terms of $\Gamma_S$, the soft anomalous dimension.
$\Gamma_S$ is calculated explicitly at one loop by evaluating the 
eikonal vertex corrections in Fig. 2.
The color basis for the hard scattering consists of only one tensor, $c=T_F$, 
and there are three eikonal lines connecting at the color vertex. 
Note that the color structure of the hard scattering for direct 
photon production is much simpler than for heavy quark or dijet production
\cite{NK,KS,KOS,NKJSRV,NKkerkyra}; 
hence, here $\Gamma_S$ is simply a $1 \times 1$ matrix
in color space.

\begin{figure}
\centerline{
\psfig{file=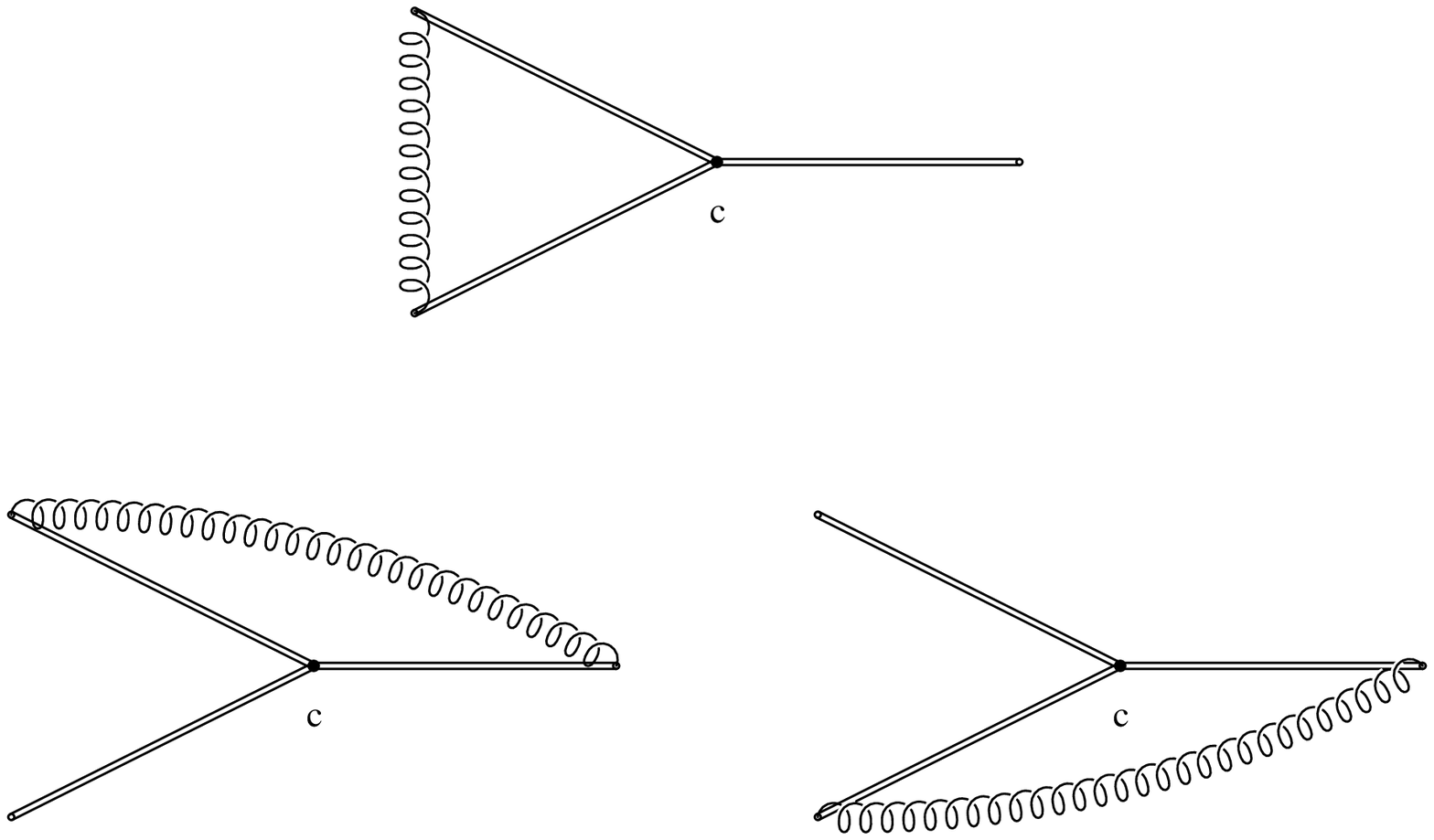,height=1.7in,width=3.2in,clip=}}
{Fig. 2. One-loop eikonal vertex corrections for 
partonic subprocesses in direct photon production; the eikonal lines 
represent the incoming partons and the outgoing jet.}
\label{fig2}
\end{figure}

The soft anomalous dimension for the process $q(p_a)+ g(p_b)  
\rightarrow \gamma(p_{\gamma}) + q(p_J)$ is given by \cite{LOS,NK,NKVD}
\beqa
\Gamma_S^{qg \rightarrow \gamma q}&=&\frac{\alpha_s}{2\pi}
\left\{C_F\left[2\ln \left(\frac{-u}{s}\right)
-\ln(4\nu_{q_a} \nu_{q_J})+2 \right] \right.
\nonumber \\ &&  \quad \quad \left.
{}+C_A\left[\ln\left(\frac{t}{u}\right)
-\ln(2 \nu_g) +1 -\pi i \right]\right\} \, .
\eeqa

The soft anomalous dimension for the process $q(p_a)+{\bar q}(p_b) 
\rightarrow  \gamma(p_{\gamma}) + g(p_J)$ is given by \cite{LOS,NK,NKVD}
\beqa
\Gamma_S^{q {\bar q} \rightarrow \gamma g}&=&\frac{\alpha_s}{2\pi}
\left\{C_F\left[-\ln(4\nu_q \nu_{\bar q})+2 -2 \pi i\right] \right.
\nonumber \\ && \quad \quad \left. 
{}+C_A\left[\ln \left(\frac{tu}{s^2}\right)
+1 -\ln(2 \nu_g) +\pi i \right]\right\} \, .
\eeqa

Note that the soft anomalous dimensions are complex and that they
also include the gauge dependent $\nu_i$ terms; 
however, all gauge dependence cancels
out explicitly in the resummed cross section.

Finally, note that the resummed cross section,         
Eq.~(\ref{rescrosec}), can be rewritten in a form \cite{NKVD} more convenient 
for the calculation of the fixed-order expansions, as
\beqa
E_{\gamma}\frac{d^3{\hat\sigma_{f_af_b\rightarrow \gamma X}}(N)}
{d^3 p_{\gamma}} &=&
H\left(\alpha_s(\mu_R^2)\right)
\exp\left[2\int_{\mu_R}^{\sqrt s} 
\frac{d\mu'}{\mu'} \beta(\alpha_s(\mu'^2))\right]
\nonumber \\ && \hspace{-40mm} \times \;
\exp \left \{ \sum_{i=a,b} \left [E^{(f_i)}(N_i)
-2\int_{\mu_F}^{2p_i \cdot \zeta}{d\mu'\over\mu'}\;
\left [\gamma_{f_i}(\alpha_s(\mu'{}^2))
-\gamma_{f_if_i}(N_i,\alpha_s(\mu'{}^2)) \right] \right] \right\}
\nonumber \\ &&  \hspace{-28mm}  \times \;
\exp \left \{E'_{(J)}(N) \right\} \;
\exp \left[2\int_{\mu_F}^{\sqrt s} {d\mu' \over \mu'}
\left(\gamma_a(\alpha_s(\mu'^2))
+\gamma_b(\alpha_s(\mu'^2))\right)\right]
\nonumber \\ && \hspace{-28mm} \times \;
{\tilde S}\left(\alpha_s(s/N^2)\right) \;
\exp \left[\int_0^1 dz \frac{z^{N-1}-1}{1-z} \,
2 {\rm Re} \Gamma_S\left(\alpha_s((1-z)^2s)\right)\right] \, .
\nonumber \\
\label{resum}
\eeqa

\mysection{NNLO expansion of the resummed cross section}

The formalism presented in the preceeding section results in 
expressions for the moments of the resummed cross section. An inverse 
Mellin transform is needed in order to obtain results for the $p_T$ and $y$ 
dependence of the cross section. Such a transform involves integrals over 
regions of phase space where the argument of the running coupling vanishes and 
one must adopt a prescription for treating the resulting singularities;
different prescriptions may give different results. On the 
other hand, it is possible to expand the moment expressions in powers of 
$\alpha_S$ and perform the inversion order-by-order at fixed $\alpha_S$. In 
this way one avoids having to adopt a specific prescription for the inversion. 
Of course, one does not have the fully resummed result after such a procedure.
However, the work of Ref. \cite{CMNOV} shows that the expansion converges 
quickly until one reaches the region very near $x_T=1.$ Therefore, in regions  
relevant to current experiments, the corrections beyond NNLO are expected 
to be small.
At any rate, this is a useful exercise for several reasons. First, one may 
check the resulting NLO ${\cal O}(\alpha \alpha_s^2)$ 
expressions with the corresponding 
terms from existing complete NLO calculations, thereby obtaining a test of the 
formalism. Second, one may generate analytical and numerical predictions 
for the dominant terms near threshold at NNLO.
These terms represent the 
leading, next-to-leading, and next-to-next-to-leading logarithms which would 
be present in a full ${\cal O}(\alpha \alpha_s^3)$ calculation.

The results can be cast in the common form given by 
\beqa
&&E_{\gamma}\frac{d^3{\hat \sigma}^{\overline{\rm MS}}_
{ij \rightarrow \gamma + X}}{d^3 p_{\gamma}}= 
\sigma^B_{ij}(s,t,u) \nonumber \\ && \hspace{-5mm}
\times \, \left\{\delta(s_4)
+\sum^2_{k=1} \left(\frac{\alpha_s(\mu_R^2)}{\pi}\right)^k
\left[c_1^k \delta(s_4) + \sum_{l=0}^{2k-1} c_{l+2}^k
\left(\frac{\ln^l (s_4/p_T^2)}{s_4}\right)_+\right]\right\} 
\label{cross_section}
\eeqa
where, for ease of notation, the subprocess labels have been suppressed on 
the $c$ coefficients. It should be noted that these coefficients depend in 
general on $s$, $t$, and $u$. The expressions for the two Born terms are 
given by 
\beq
\sigma^B_{qg}= -\frac{1}{N_c}\frac{\alpha \alpha_s}{s} e_q^2 
\left(\frac{s}{t}+\frac{t}{s}\right)
\eeq
and
\beq
\sigma^B_{q\overline q}=\frac{2 C_F}{N_c}\frac{\alpha \alpha_s}{s} e_q^2
\left(\frac{t}{u}+\frac{u}{t}\right)
\eeq
where $e_q$ is the charge of a quark of type $q$. The 
${\cal O}(\alpha \alpha_s^2)$ results for the coefficients $c$ are given by 
\beqa
c_3^1(qg)&=& C_F+2C_A \nonumber \\
c_2^1(qg)&=&-\frac{3}{4}C_F-(C_F+C_A)\ln\left(\frac{\mu_F^2}{p_T^2}\right) 
\nonumber \\
c_1^1(qg)&=&\left(-\frac{\beta_0}{4}-\frac{3}{4}C_F-C_F\ln\left(\frac{-t}{s}
\right)-C_A\ln\left(\frac{-u}{s}\right)\right) \ln\left(\frac{\mu_F^2}{p_T^2}
\right) \nonumber \\
&+&\frac{\beta_0}{4}\ln\left(\frac{\mu_R^2}{p_T^2}\right)
\eeqa
and
\beqa
c_3^1(q\overline q)&=& 4 C_F-C_A \nonumber \\
c_2^1(q\overline q)&=& -\frac{\beta_0}{4}-2C_F\ln\left(\frac{\mu_F^2}{p_T^2}
\right) \nonumber \\
c_1^1(q\overline q)&=&C_F\left(-\frac{3}{2}-\ln\left(\frac{p_T^2}{s}\right)
\right)\ln\left(\frac{\mu_F^2}{p_T^2}\right)+\frac{\beta_0}{4}
\ln\left(\frac{\mu_R^2}{p_T^2}\right).
\eeqa
These expressions are equivalent to the corresponding terms in the exact NLO 
results quoted in Ref.~\cite{GV}. The comparison is most easily made by 
rewriting the results in terms of the variables $v$ and $w$ used in 
\cite {GV} as is shown in the Appendix. Note that, with respect to the 
$\delta(s_4)$ contribution, the expansion reproduces only
the scale-dependent terms in $c_1^1$. The full $\delta(s_4)$ 
terms can be simply read off from the exact NLO calculation.

The NNLO corrections in the $\overline {\rm MS}$ scheme for the 
$q g \rightarrow \gamma q$ subprocess are given by the following coefficients: 
\beqa   
c_5^2(qg)&=&\frac{1}{2}\left(C_F+2C_A\right)^2 \nonumber \\
c_4^2(qg)&=&-\frac{3}{2} (C_F+2C_A)\left(\frac{3}{4}C_F
+(C_F+C_A)\ln\left(\frac{\mu_F^2}{p_T^2}\right)\right) 
-\frac{\beta_0}{2}\left(\frac{C_F}{4}+C_A\right) \nonumber \\
c_3^2(qg)&=&(C_F+2C_A)\left[c_a'^{qg} \ln\left(\frac{\mu_F^2}{p_T^2}\right)
+c_1'^{qg}\right] 
+\left[\frac{3}{4}C_F
+(C_F+C_A)\ln\left(\frac{\mu_F^2}{p_T^2}\right)\right]^2 
\nonumber \\ 
&+&(C_F+2C_A) \frac{\beta_0}{4} \ln \left(\frac{\mu_R^2}{p_T^2}\right)
+\frac{1}{2} K (C_F+2C_A)
+\frac{3}{16}\beta_0 C_F \nonumber \\
&-&\frac{\pi^2}{6}(C_F+2C_A)^2
\eeqa
with 
\beq 
c_a'^{qg}=-\frac{\beta_0}{4}-\frac{3}{4}C_F
-C_F \ln\left(\frac{-t}{s}\right)
-C_A \ln\left(\frac{-u}{s}\right)
\eeq
and 
\beqa 
c_1'^{qg}&=&\frac{\beta_0}{4}\ln\left(\frac{\mu_R^2}{p_T^2}\right)
-\frac{7}{4}C_F-\frac{5}{4}N_c\ln^2\left(\frac{s+t}{s}\right)
\nonumber \\ && 
{}-\frac{1}{2}C_F\ln^2\left(\frac{-t}{s}\right)
+\frac{3}{4}C_F\ln\left(\frac{s+t}{s}\right)
-\frac{3}{2}C_F\ln\left(\frac{p_T^2}{s}\right)
\nonumber \\ && \hspace{-15mm}
+\frac{N_c}{T_{qg}}\left[-\frac{t}{s}\ln^2\left(\frac{s+t}{s}\right)
-\frac{t}{2s}\ln\left(\frac{-t}{s}\right)
\right.
\nonumber \\ && 
{}+\frac{t}{s}\ln\left(\frac{s+t}{s}\right)
+\frac{\pi^2}{4}\left(\frac{-t}{s}\right)\left(\frac{2s+t}{s}\right)
-\frac{t}{4s}\left(\frac{2s+t}{s}\right)\ln^2\left(\frac{-t}{s}\right)
\nonumber \\ && \left. 
{}+\frac{t}{2s}\left(\frac{2s+t}{s}\right)\ln\left(\frac{-t}{s}\right)
\ln\left(\frac{s+t}{s}\right)\right]
\nonumber \\ && \hspace{-15mm}
{}+\frac{C_F}{T_{qg}}\left[\frac{1}{2}\left(\frac{3s+t}{s}\right)
\ln\left(\frac{-t}{s}\right)
+\frac{\pi^2}{6}\left(1-4\frac{(s+t)}{s}+5\frac{(s+t)^2}{s^2}\right)\right.
\nonumber \\ && \hspace{-5mm}
{}-\frac{1}{4}\left(3\frac{(s+t)^2}{s^2}+\frac{2t}{s}\right)
\ln\left(\frac{s+t}{s}\right)
+\frac{1}{2}\left(3\frac{(s+t)^2}{s^2}-\frac{2t}{s}\right)
\ln^2\left(\frac{s+t}{s}\right)
\nonumber \\ && \hspace{-10mm} \left.
{}+\frac{1}{2} \left(\frac{(s+t)^2}{s^2}+\frac{t^2}{s^2}\right)
\ln^2\left(\frac{-t}{s}\right)
-\left(\frac{(s+t)^2}{s^2}+\frac{t^2}{s^2}\right)
\ln\left(\frac{-t}{s}\right)\ln\left(\frac{s+t}{s}\right)\right] \, .
\nonumber \\
\eeqa

The corresponding coefficients for the $q \overline q \rightarrow \gamma g$ 
subprocess are:
\beqa
c_5^2(q\overline q)&=&\frac{1}{2}\left(4C_F-C_A\right)^2 \nonumber \\
c_4^2(q\overline q)&=&-3\left(4C_F-C_A\right)C_F\ln\left(\frac{\mu_F^2}{p_T^2}
\right) - \frac{\beta_0}{2}\left(5C_F-\frac{3}{2}C_A\right) \nonumber \\
c_3^2(q\overline q)&=&(4C_F-C_A)\left[c_a'^{q \bar q}\ln\left(\frac{\mu_F^2}
{p_T^2}\right)+c_1'^{q \bar q}\right] 
+\left[\frac{\beta_0}{4}+2C_F\ln \left(\frac{\mu_F^2}{p_T^2}\right)\right]^2
\nonumber \\ && \hspace{-20mm} 
+(4C_F-C_A)\frac{\beta_0}{4}\ln\left(\frac{\mu_R^2}{p_T^2}\right)
+(4C_F-C_A)\frac{K}{2}+\frac{\beta_0^2}{16}
-\frac{\pi^2}{6}(4C_F-C_A)^2 
\eeqa
with 
\beq 
c_a'^{q \bar q}=C_F\left[-\frac{3}{2}-\ln\left(\frac{p_T^2}{s}\right)\right]
\eeq
and 
\beqa 
c_1'^{q \bar q}&=&\frac{\beta_0}{4}\ln\left(\frac{\mu_R^2}{p_T^2}\right)
+\frac{1}{2}K-\frac{7}{2}C_F+\frac{\pi^2}{2}\left(C_F-\frac{C_A}{6}\right)
\nonumber \\ &&
{}+C_F\ln\left(\frac{-t}{s}\right)\ln\left(\frac{-u}{s}\right)
-C_F\ln^2\left(\frac{p_T^2}{s}\right)
-\frac{3}{2}C_F\ln\left(\frac{p_T^2}{s}\right)
\nonumber \\ && \hspace{-10mm}
{}+\frac{C_F}{2T_{q \bar q}}\left[-\frac{2u}{s}\ln\left(\frac{-t}{s}\right) 
-\frac{2t}{s}\ln\left(\frac{-u}{s}\right)
+\frac{u^2}{s^2}\ln\left(\frac{-t}{s}\right)
+\frac{t^2}{s^2}\ln\left(\frac{-u}{s}\right)
\right.
\nonumber \\ && \left.
{}+\left(\frac{3u^2}{s^2}-\frac{2t}{s}\right)\ln^2\left(\frac{-u}{s}\right)
+\left(\frac{3t^2}{s^2}-\frac{2u}{s}\right)\ln^2\left(\frac{-t}{s}\right)
\right]
\nonumber \\ && \hspace{-10mm}
-\frac{N_c}{2 T_{q \bar q}}\left[\frac{tu}{s^2}\ln\left(\frac{tu}{s^2}\right)
-\frac{(t^2+2ts)}{2s^2}\ln^2\left(\frac{-t}{s}\right)
-\frac{(u^2+2us)}{2s^2}\ln^2\left(\frac{-u}{s}\right) \right] \, .
\nonumber \\
\eeqa
In the above expressions for $c_1^{'qg}$ and $c_1^{'q\overline q}$, $T_{qg}$ 
and $T_{q\overline q}$ are given by 
\beqa
T_{qg}&=&(s^2+t^2)/s^2 \nonumber \\
T_{q\overline q}&=&(t^2+u^2)/s^2 \, .
\eeqa
Note that, at NNLO, all the NNLL terms, i.e. 
${\cal O}\left(\left[\ln(s_4/p_T^2)/s_4\right]_+\right)$ terms,
have been derived by matching with the exact NLO cross section. The 
scale-dependent terms at that accuracy can also be derived by the 
straightforward expansion of the resummed cross section. 
The result is the same, which provides a nice cross check.
As an additional check, an explicit variation of the hadronic cross
section with respect to scale yields zero up to $(1/s_4)_+$ terms.
From the expansion of the resummed cross section we can also derive 
the $(1/s_4)_+$ terms containing squared logarithms of the scale.
These terms in the coefficient $c_2^2(qg)$ are
\beqa
&& \ln^2{\left(\frac{\mu_F^2}{p_T^2}\right)}\, \left(C_F + C_A\right)
\left[C_F \left(\ln\left(\frac{-t}{s}\right) +\frac{3}{4}\right)
+C_A \ln\left(\frac{-u}{s}\right)
+\frac{3\beta_0}{8} \right] 
\nonumber \\ &&  
{}-\frac{\beta_0}{2} \, (C_F+C_A) \, \ln\left(\frac{\mu_R^2}{p_T^2}\right) \, 
\ln\left(\frac{\mu_F^2}{p_T^2}\right) \, ,
\eeqa
while the corresponding terms in the coefficient $c_2^2(q {\bar q})$ are
\beq
\ln^2\left(\frac{\mu_F^2}{p_T^2}\right) \,
C_F \left[C_F\left(3+2\ln\left(\frac{p_T^2}{s}\right)\right)
+\frac{\beta_0}{4}\right]   
-\beta_0 \, C_F \, \ln\left(\frac{\mu_R^2}{p_T^2}\right) \, 
\ln\left(\frac{\mu_F^2}{p_T^2}\right) \, .
\eeq
Since we don't reproduce the full $c_2^2$ coefficients, we don't include them
in our numerical results in the next section. 

Finally, we note that all gauge dependence cancels explicitly in
the NNLO expansions, as expected.

\mysection{Numerical results}

In order to examine the phenomenological consequences of the NNLO expressions 
presented in this paper, the ${\cal O} (\alpha \alpha_s^2)$ program of 
Ref. \cite{BOO} is used as a starting point. This includes the lowest 
order $2 \rightarrow 2$ subprocesses and their virtual corrections, 
the ${\cal O}(\alpha \alpha_s^2)$ 
$2 \rightarrow 3$ subprocesses, and a fragmentation contribution calculated 
by convoluting $2 \rightarrow 2$ parton-parton scattering subprocesses with 
appropriate photon fragmentation functions. The NNLO expressions are then 
added to the output of this program. It is important to note that the 
fragmentation contribution has not been modified by the inclusion of any 
resummed terms. On the other hand, for fixed target energies the fragmentation 
component is generally small in the $p_T$ region spanned by the data. 
The results discussed in this section have been obtained with the  
renormalization and factorization scales set equal to each other. In each 
case two scale choices have been used, $p_T/2$ and $2p_T$. The CTEQ5M 
\cite{CTEQ5} parton distributions have been used. In each of the comparisons 
presented below, the calculation for the invariant cross section has been 
averaged over the appropriate region of rapidity or transverse momentum as 
specified for the data set being shown.

\begin{figure}
\centerline{
\psfig{file=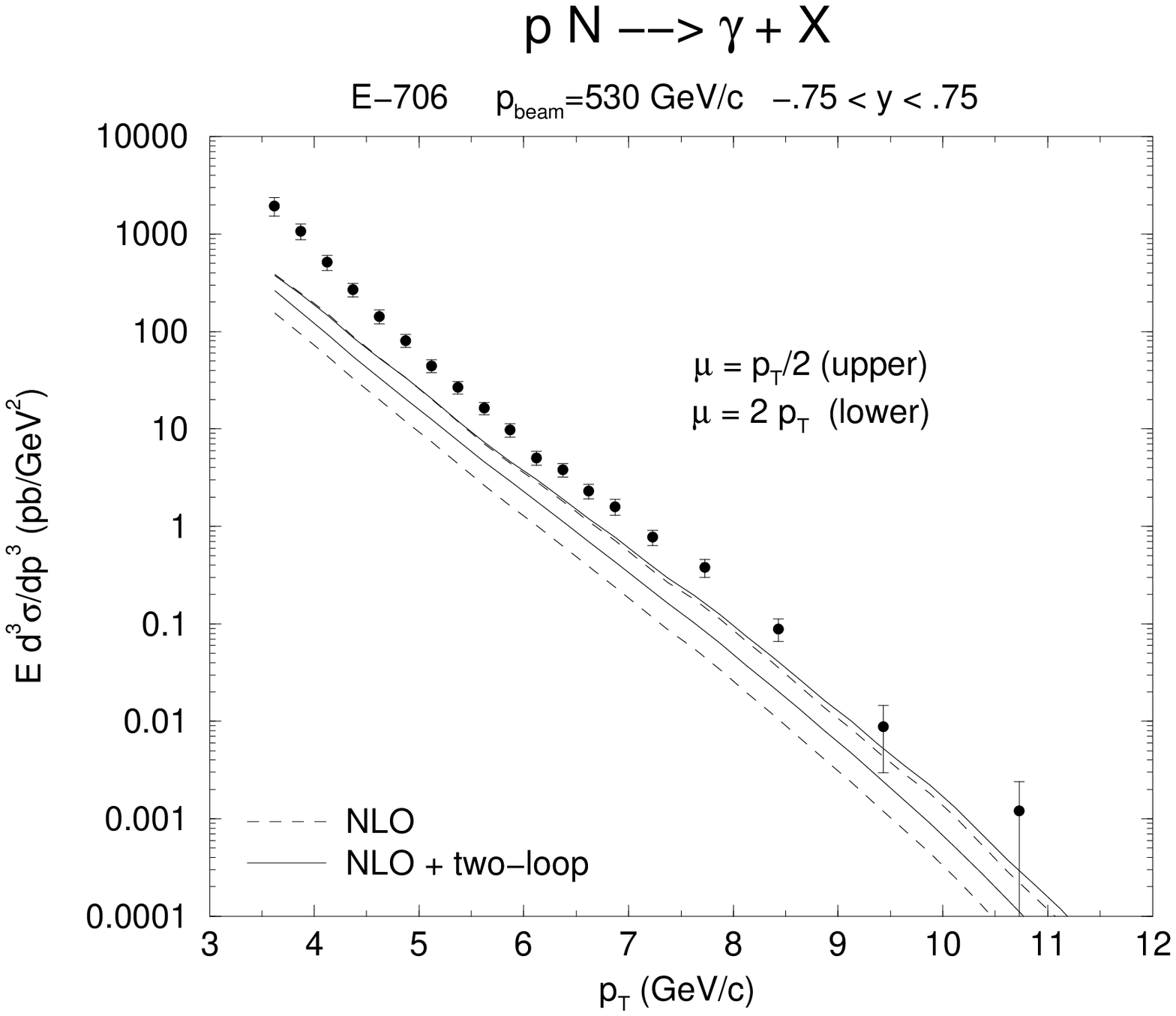,height=3in,width=4in,clip=}}
{Figure 3. NLO and NNLO results for direct photon production
in hadronic collisions compared to data from the E-706 Collaboration 
\cite{E706} at p$_{\rm beam}$=530 GeV/c.}
\label{fig3}
\end{figure}

The curves in Fig. 3 show the NLO and NNLO predictions for the invariant 
cross section for direct photon production using a proton beam on a beryllium 
target. The predictions are compared with the experimental results at 
$p_{\rm beam}=530 \;{\rm GeV/c}$ from the E-706 Collaboration \cite{E706}. The 
theoretical curves have been multipled by a nuclear correction factor of 
1.09 \cite{E706}. Comparing the band formed by the solid curves with that 
formed by the dashed curves shows that the inclusion of the NNLO contributions 
reduces the scale dependence by about a factor of two. A similar reduction 
has been noted in \cite{CMNOV}. The origin of this reduction has been 
discussed in both \cite{CMNOV} and \cite{EPIC99}. Furthermore, over the $p_T$ 
range shown in Fig. 3 the NNLO corrections for $\mu=p_T/2$ are rather small. 
Thus, the inclusion of the threshold resummation corrections is unable to 
bring the predictions into agreement with these data. As noted in the 
introduction, it has been argued that initial state radiation can lead to 
non-zero values for the transverse momentum of the colliding partons and that 
the inclusion of such effects may bring the prediction into agreement with 
the data. 
However, to date this has only been treated in a phenomenological manner 
\cite{kT} and a more definitive theoretical treatment is needed before 
firm conclusions can be drawn. 
Some recent work in this direction can be found in \cite{Li}.

\begin{figure}
\centerline{
\psfig{file=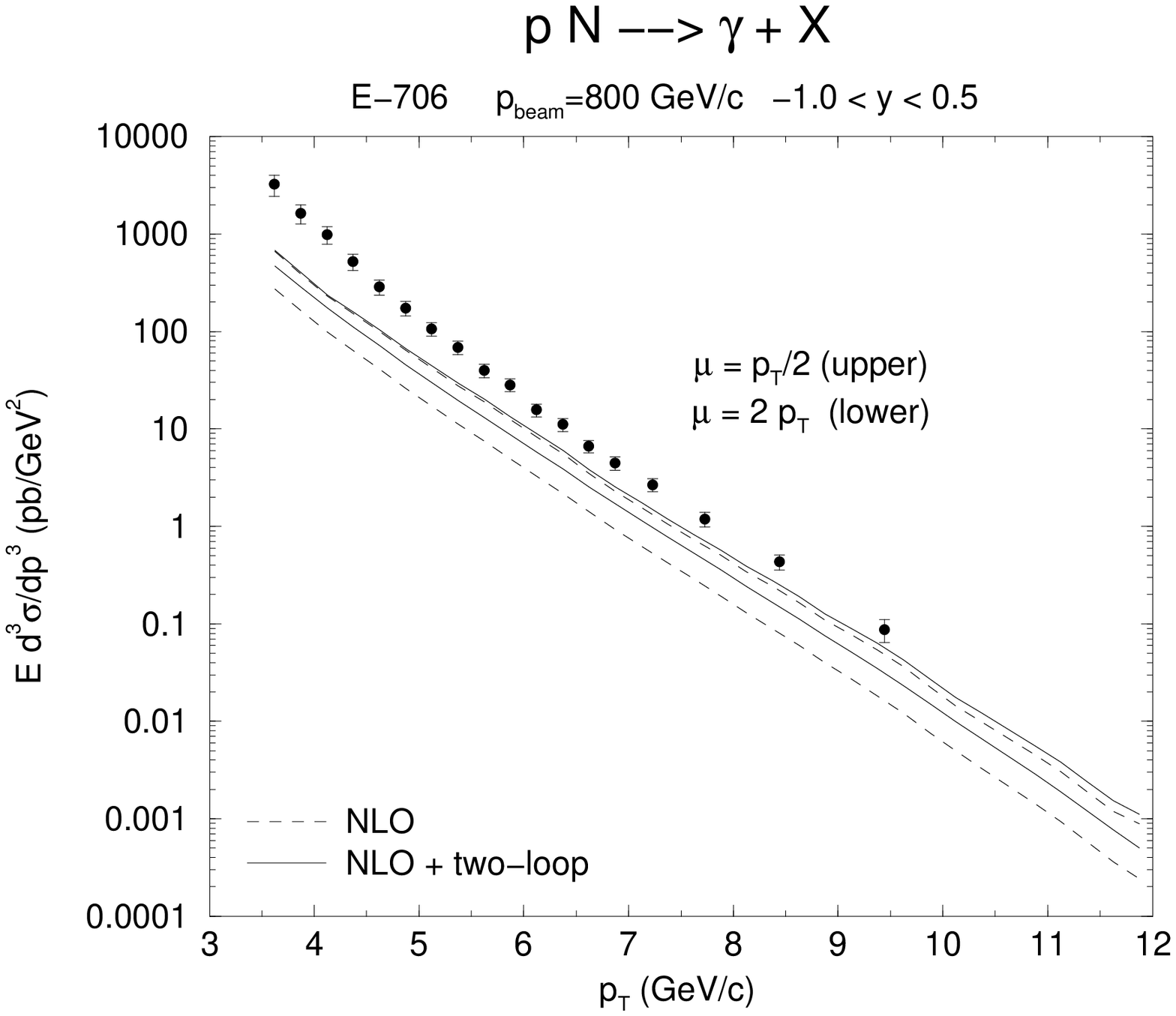,height=3in,width=4in,clip=}}
{Figure 4. NLO and NNLO results for direct photon production
in hadronic collisions compared to data from the E-706 Collaboration 
\cite{E706} at 
p$_{\rm beam}$=800 GeV/c.}
\label{fig4}
\end{figure}

A similar pattern of behavior is shown in Fig. 4 for the 800 GeV/c results 
from the E-706 Collaboration. Again, the NNLO corrections are not 
sufficient to bring the predictions into agreement with the data. However, 
the reduction in the overall scale dependence remains.

\begin{figure}
\centerline{
\psfig{file=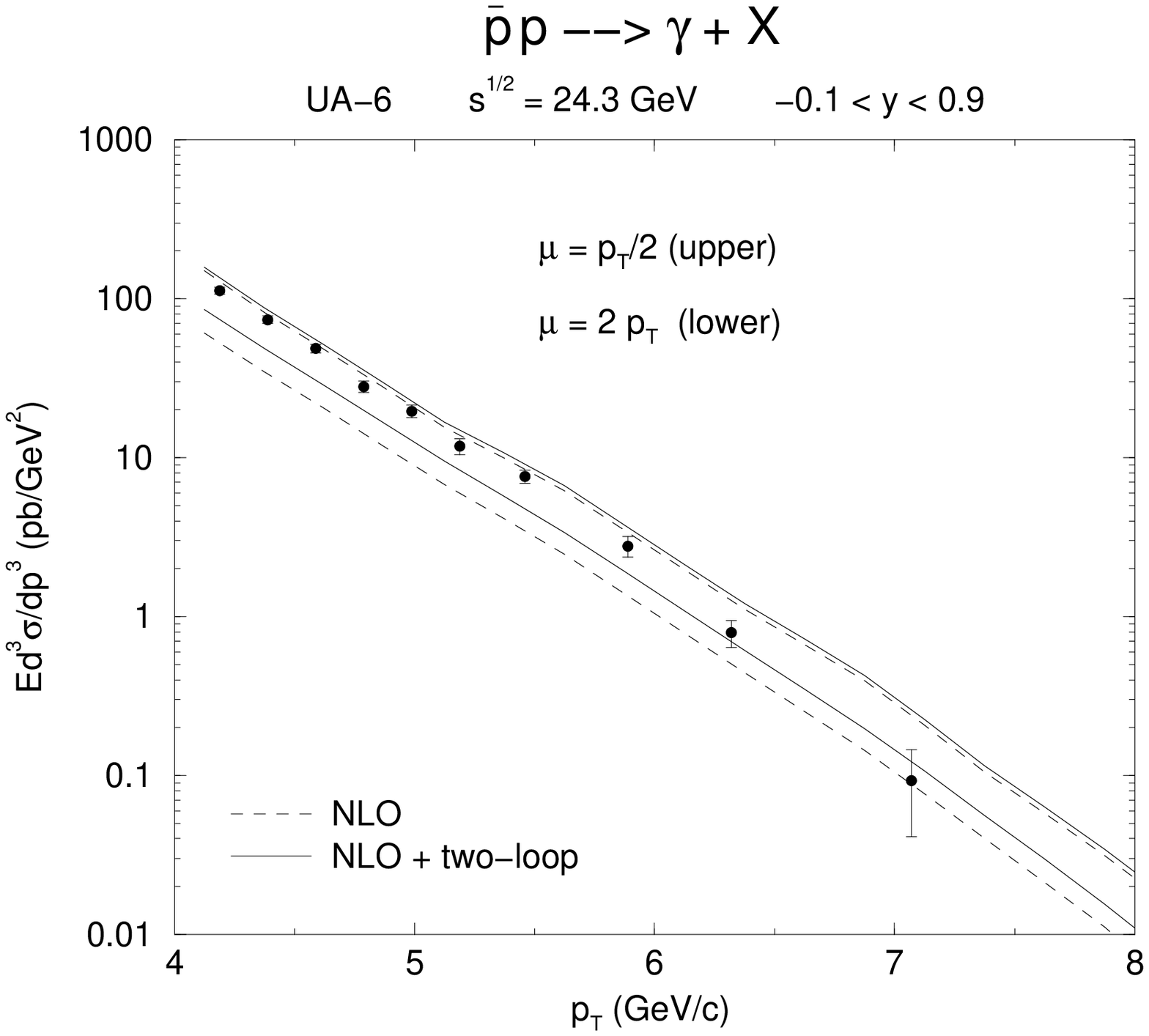,height=3in,width=4in,clip=}}
{Figure 5. NLO and NNLO results for direct photon production
in hadronic collisions compared to $p \overline p$ data from the 
UA-6 Collaboration \cite{UA6}.}
\label{fig5}
\end{figure}

The UA-6 Collaboration has published data \cite{UA6} for direct photon 
production in both $p\overline p$ and $pp$ collisions. Results are available 
for both the rapidity and transverse momentum dependence of the invariant 
cross section. In Fig. 5 the results for the $p \overline p$ collisions are 
shown. In this case the band formed by the NLO predictions overlaps the 
data at all values of $p_T$ shown. The reduction in scale dependence provided 
by the inclusion of the NNLO terms is also evident. 

\begin{figure}
\centerline{
\psfig{file=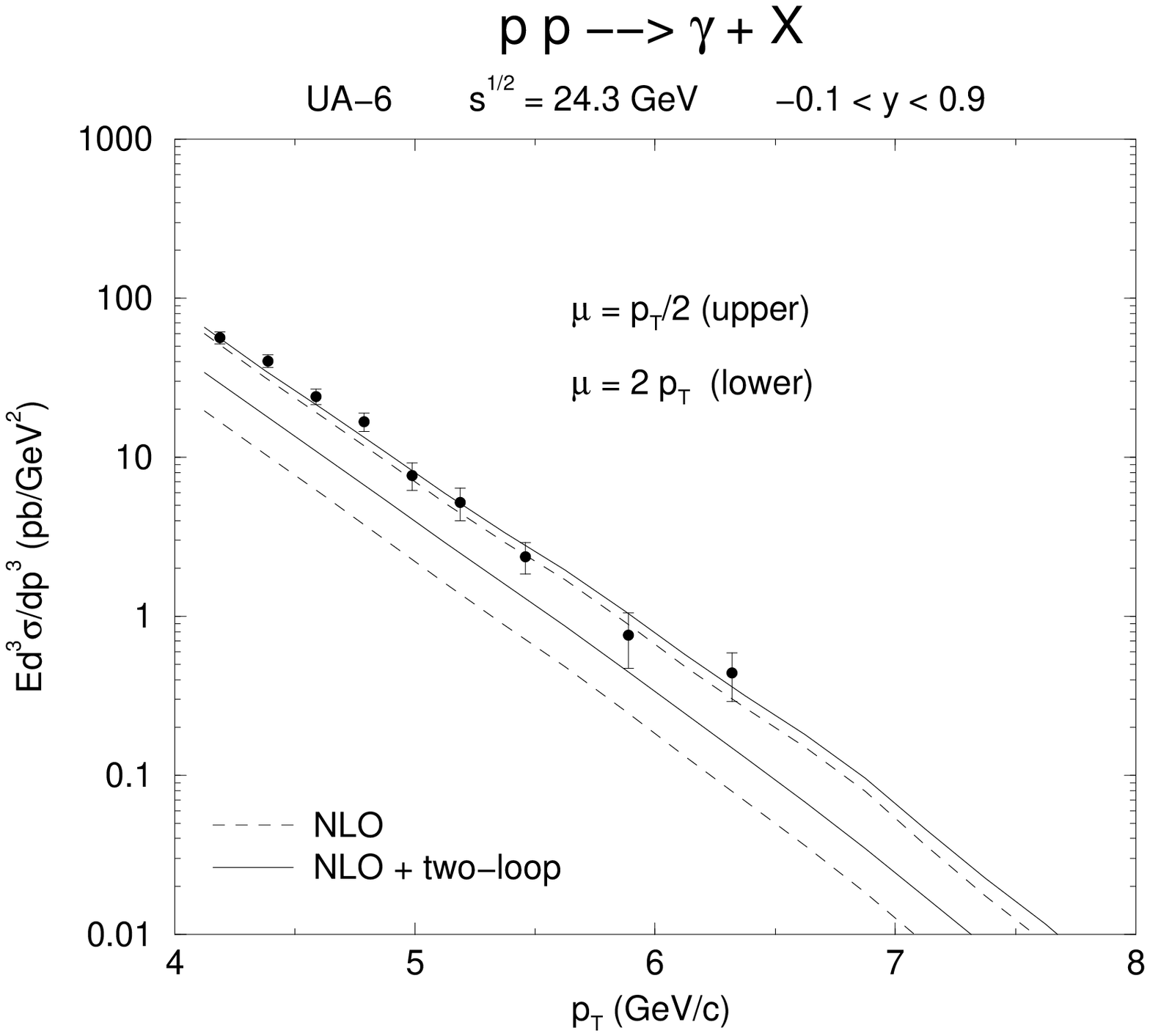,height=3in,width=4in,clip=}}
{Figure 6. NLO and NNLO results for direct photon production
in hadronic collisions compared to $pp$ data from the UA-6 Collaboration 
\cite{UA6}.}
\label{fig6}
\end{figure}

In Fig. 6 the corresponding results for $pp$ collisions are compared with the 
UA-6 data. One can see that the predictions with $\mu=p_T/2$ (with or 
without the NNLO terms) are compatible with the data and the now familiar 
reduction in scale dependence is evident.

\begin{figure}
\centerline{
\psfig{file=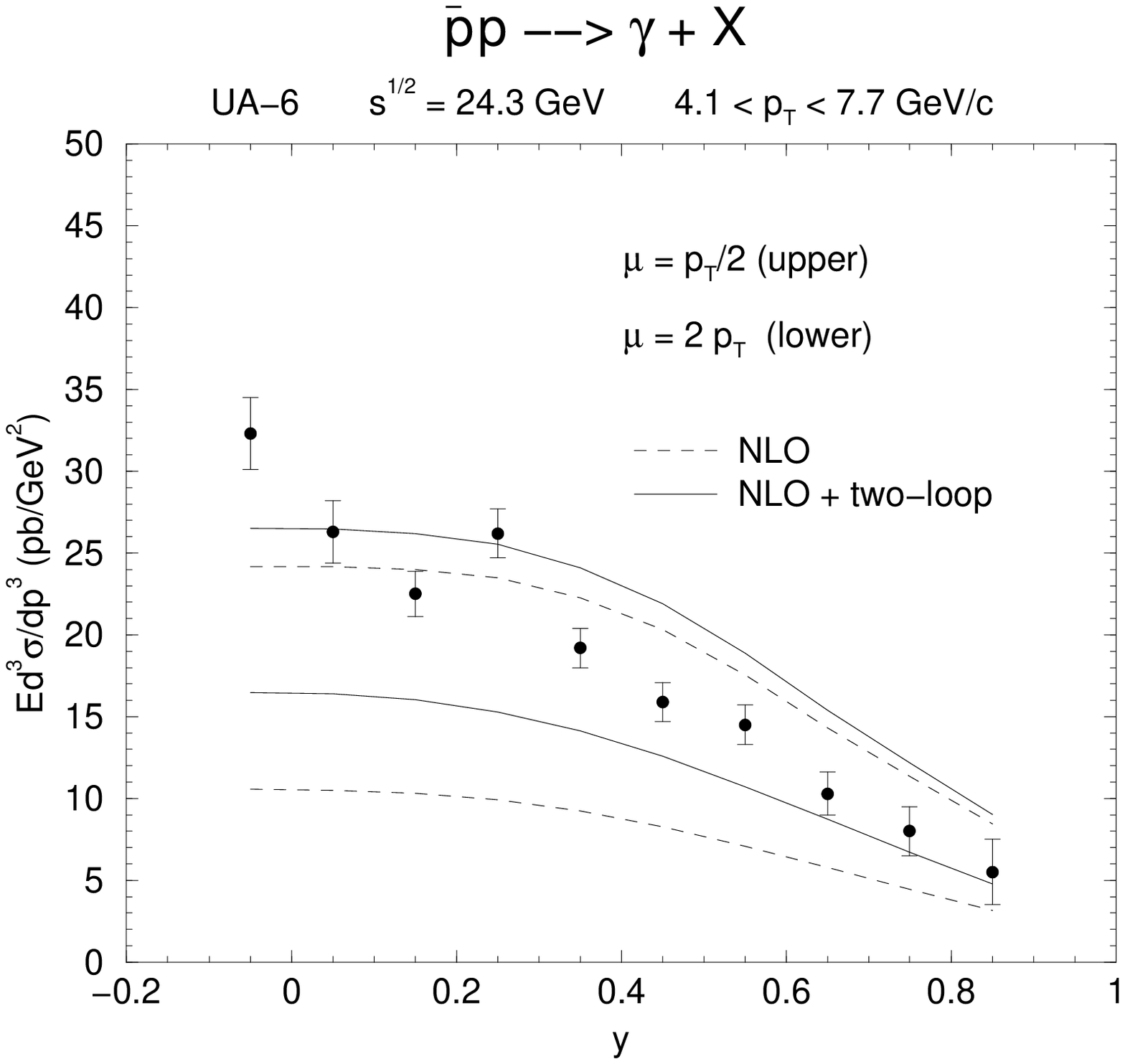,height=3in,width=4in,clip=}}
{Figure 7. NLO and NNLO results for the rapidity distribution for 
direct photon production in hadronic collisions compared to $p \overline p$ 
data from the 
UA-6 Collaboration \cite{UA6}.}
\label{fig7}
\end{figure}

The resummation formalism presented in this paper allows one to study the 
dependence on both the transverse momentum and the rapidity of the photon. 
In Figs. 7 and 8 the predictions are compared to the rapidity distributions 
published by the UA-6 Collaboration \cite{UA6} for $p\overline p$ and 
$pp$ collisions, 
respectively. Comparing the NLO and NNLO results shows that the rapidity 
dependence of the relative size of the NNLO corrections is mild. In each case 
the reduced scale dependence is evident.

\begin{figure}
\centerline{
\psfig{file=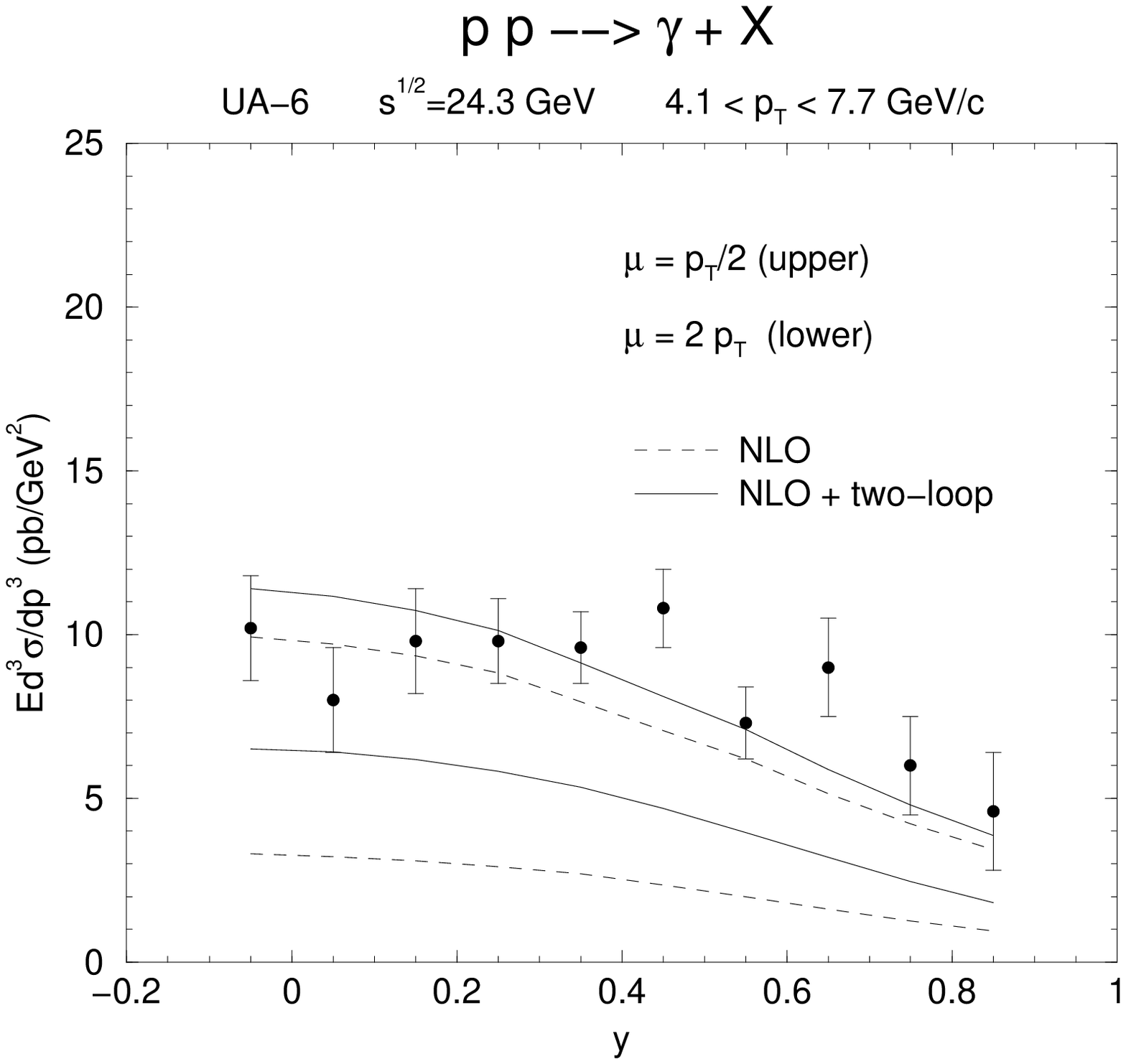,height=3in,width=4in,clip=}}
{Figure 8. NLO and NNLO results for the rapidity distribution for direct 
photon production in hadronic collisions compared to $pp$ data from the UA-6 
Collaboration \cite{UA6}.}
\label{fig8}
\end{figure}

\mysection{Conclusion}

In this paper the application of threshold resummation to direct photon 
production has been studied using the formalism of Refs. \cite{LOS,NK}. The 
expressions for the moments of the resummed cross section have been expanded 
to ${\cal O}(\alpha \alpha_S^3)$ and expressions for the leading,
next-to-leading, 
and next-to-next-to-leading logarithmic contributions have been given. The 
NNLL estimates for the ${\cal O} (\alpha \alpha_S^3)$ contributions have been 
incorporated into an existing next-to-leading logarithm 
${\cal O} (\alpha \alpha_S^2)$ program and compared 
to high statistics fixed target data. The scale dependence of the 
results is markedly decreased by the inclusion of the NNLO terms. 
Furthermore, it was found that the size of the corrections is small when the 
factorization and renormalization scales are chosen to be $\mu=p_T/2$. 
Nevertheless, even with the inclusion of these threshold corrections, 
it is still not possible to achieve a good description of all of the fixed 
target data. Additional study of this process will be required, both 
theoretically and experimentally.

\mysection*{Acknowledgements}

The authors wish to thank Vittorio Del Duca, Eric Laenen, and George Sterman 
for many useful discussions. This work was supported in part by the U.S. 
Department of Energy.

\appendix
\section{Alternative form of the NNLO expansion}
\label{app-exp}

In this Appendix we rewrite our NLO and NNLO expansions in terms
of the variables $v=(s+t)/s$ and $w=-u/(s+t)$ used in Ref. \cite{GV}.  

The factorized form of the cross section for direct photon 
production can be rewritten as:
\beqa
E_{\gamma}\frac{d^3\sigma_{h_Ah_B\rightarrow\gamma X}}{d^3 p_{\gamma}}&=&
\sum_{f}\int dx_a dx_b \,  \phi_{f_a/h_A}(x_a,\mu_F^2) \,
\phi_{f_b/h_B}(x_b,\mu_F^2) 
\nonumber \\ && \hspace{5mm}
\times \; \frac{1}{\pi s v} \frac{d{\hat \sigma}_{f_a f_b\rightarrow \gamma X}}
{dv \, dw} (w,s,v,\mu_F,\alpha_s(\mu_R^2))  \, .
\label{factalt}
\eeqa
The threshold region is given by $w=1$.
In general, $\hat{\sigma}$ includes distributions with respect
to $1-w$ at $n$th order in $\alpha_s$ of the type
\beq
\left[\frac{\ln^{m}(1-w)}{1-w} \right]_+, \hspace{10mm} m\le 2n-1\, .
\label{wplus}
\eeq

Before presenting the NNLO expansions of the resummed cross section 
in terms of the variables $v$ and $w$, 
we want to emphasize that crossing $t$ and $u$ mixes singular
and non-singular terms in $1-w$. The non-singular terms are required
in order to have the correct symmetrization under $t \leftrightarrow u$
away from threshold.  
Therefore in producing numerical results we use the expansions
in terms of the variable $s_4$ as presented in Section 3, where the symmetry 
is manifest and there is no
mixing between singular and non-singular terms.

It is convenient to express the results in a form similar to that in 
Eq. (\ref{cross_section}):
\beqa   
&&vw(1-v)s\frac{d{\hat \sigma}^{\overline{\rm MS}}_{i j \rightarrow 
\gamma +X}}{dv \, dw} 
=\tilde \sigma^B_{i j}(s,v) 
\nonumber \\ && \hspace{-5mm} \times \, 
\left\{\delta(1-w)+\sum_{k=1}^2\left(\frac{\alpha_s(\mu_R^2)}{\pi}\right)^k
\left[\tilde c_1^k \delta(1-w) + \sum_{l=0}^{2k-1}\tilde c_{l+2}^k 
\left(\frac{\ln^l(1-w)}{1-w}\right)_{+}\right]\right\}\, .
\nonumber \\
\eeqa

The expressions for the two Born terms are 
\beq
\tilde \sigma^B_{qg}(s,v)=\frac{1}{N_c}\pi\alpha\alpha_se_q^2T_{qg}v
\eeq
and
\beq
\tilde \sigma^B_{q\overline q}(s,v)=\frac{2C_F}{N_c}\pi\alpha\alpha_se_q^2
T_{q\overline q}
\eeq
where $T_{qg}=1+(1-v)^2$ and $T_{q \overline q}=v^2+(1-v)^2$.

The coefficients for the NLO terms are as follows:
\beqa
\tilde c_3^1(qg)&=&C_F+2C_A \nonumber \\
\tilde c_2^1(qg)&=&C_F\left(-\frac{3}{4}+\ln v\right)-C_A\ln\left(\frac{1-v}{v}\right)
-(C_F+C_A)\ln\left(\frac{\mu_F^2}{s}\right) \nonumber \\
\tilde c_1^1(qg)&=& \left(-\frac{\beta_0}{4}-\frac{3}{4}C_F
+C_A\ln\left(\frac{1-v}{v}\right)\right)
\ln\left(\frac{\mu_F^2}{s}\right) 
+\frac{\beta_0}{4}\ln\left(\frac{\mu_R^2}{s}\right)
\nonumber \\
\eeqa
and
\beqa
\tilde c_3^1(q\overline q)&=&(4C_F-C_A) \nonumber \\  
\tilde c_2^1(q\overline q)&=&-2C_F\ln\left(\frac{1-v}{v}\right) 
+C_A \ln(1-v) -\frac{\beta_0}{4}
-2 C_F \ln\left(\frac{\mu_F^2}{s}\right) \nonumber \\
\tilde c_1^1(q \overline q)&=&C_F \left(-\frac{3}{2}
+\ln\left(\frac{1-v}{v}\right)\right)
\ln\left(\frac{\mu_F^2}{s}\right) 
+\frac{\beta_0}{4}\ln\left(\frac{\mu_R^2}{s}\right)\, .
\eeqa

These results are in full agreement with the exact NLO results in 
Ref.~\cite{GV}.  
Note that the expansion reproduces only the scale-dependent 
terms in ${\tilde c}_1^1$. 
The full $\delta(1-w)$ terms can be simply read off
from the exact NLO calculation.

The coefficients for the NNLO corrections for the $qg \rightarrow \gamma q$
subprocess are:
\beqa 
\tilde c_5^2(qg)&=& \frac{1}{2} (C_F+2C_A)^2 \nonumber \\
\tilde c_4^2(qg)&=& \frac{3}{2} (C_F+2C_A)
\left[C_F\left(-\frac{3}{4}+\ln v\right)-C_A\ln\left(\frac{1-v}{v}\right)
\right.
\nonumber \\ && \left. \hspace{28mm}
{}-(C_F+C_A)\ln\left(\frac{\mu_F^2}{s}\right)\right] 
-\frac{\beta_0}{2}\left(\frac{C_F}{4}+C_A\right) 
\nonumber \\
\tilde c_3^2(qg)&=&
(C_F+2C_A)\left[c_a^{qg} \ln\left(\frac{\mu_F^2}{s}\right)
+c_1^{qg}\right] 
\nonumber \\ 
&+&\left[C_F\left(-\frac{3}{4}+\ln v \right)
-C_A \ln \left(\frac{1-v}{v}\right)
-(C_F+C_A)\ln\left(\frac{\mu_F^2}{s}\right)\right]^2 
\nonumber \\ 
&+&(C_F+2C_A) \frac{\beta_0}{4} \ln \left(\frac{\mu_R^2}{s}\right)
+\frac{1}{2} K (C_F+2C_A)
\nonumber \\ 
&+&\beta_0\left[\frac{C_F}{4}\left(\frac{3}{4}-\ln v\right)
+\frac{C_A}{2}\ln \left(\frac{1-v}{v}\right)\right]\nonumber \\
&-&\frac{\pi^2}{6}(C_F+2C_A)^2 \, ,
\eeqa
with \cite{GV}
\beq 
c_a^{qg}=-\frac{\beta_0}{4}-\frac{3}{4}C_F+C_A \ln\left(\frac{1-v}{v}\right)
\eeq
and 
\beqa 
c_1^{qg}&=&\frac{\beta_0}{4}\ln\left(\frac{\mu_R^2}{s}\right)-\frac{7}{4}C_F
\nonumber \\ && \hspace{-15mm}
{}+\frac{N_c}{T_{qg}}\left[-\frac{1}{4}(v^2-2(1-v))\ln^2v
+\frac{1}{2}(1-v) \, \ln(1-v)-(1-v)\ln v \right.
\nonumber \\ && \left.
{}+\frac{1}{4}\pi^2(1-v^2)
+\frac{1}{4}(1-v^2)\ln^2(1-v) 
-\frac{1}{2}(1-v^2)\ln v \, \ln(1-v)\right]
\nonumber \\ && \hspace{-15mm}
{}+\frac{C_F}{T_{qg}}\left[\frac{1}{2}(1+2v)\ln(1-v)
+\frac{\pi^2}{6}(1-4v+5v^2)\right.
\nonumber \\ && 
{}-\frac{1}{4}(3v^2-2(1-v))\ln v 
+\frac{1}{2}\left(3v^2+2(1-v)\right)\ln^2 v 
\nonumber \\ && \left.
{}+\frac{1}{2} \left(v^2+(1-v)^2\right) \ln^2 (1-v)
-\left(v^2+(1-v)^2\right) \ln v \, \ln (1-v)\right] \, .
\nonumber \\
\eeqa

The corresponding NNLO terms for the subprocess $q {\bar q} \rightarrow 
\gamma g$ are:
\beqa
\tilde c_5^2(q\overline q)&=&\frac{1}{2}(4C_F-C_A)^2 \nonumber \\
\tilde c_4^2(q\overline q)&=& \frac{3}{2}(4C_F-C_A) \left[-2C_F
\ln\left(\frac{1-v}{v}\right)+C_A \ln(1-v) \right.
\nonumber \\ && \left. \hspace{28mm}
{}-2 C_F\ln\left(\frac{\mu_F^2}{s}\right)\right]   
-\frac{\beta_0}{2}\left(5C_F-\frac{3}{2}C_A\right) 
\nonumber \\
\tilde c_3^2(q \overline q)&=&(4C_F-C_A)\left[c_a^{q \bar q} 
\ln\left(\frac{\mu_F^2}{s}\right)
+c_1^{q \bar q}\right] 
\nonumber \\ 
&+&\left[-2C_F\ln \left(\frac{1-v}{v}\right)+C_A \ln(1-v)
-\frac{\beta_0}{4}-2C_F\ln \left(\frac{\mu_F^2}{s}\right)\right]^2
\nonumber \\ 
&+&(4C_F-C_A)\frac{\beta_0}{4}\ln\left(\frac{\mu_R^2}{s}\right)
+(4C_F-C_A)\frac{K}{2}+\frac{\beta_0^2}{16}
\nonumber \\ 
&+&\beta_0\left[C_F\ln\left(\frac{1-v}{v}\right)
+\frac{C_A}{4}\left(\ln v-2\ln(1-v)\right)\right]
\nonumber \\ 
&-&\frac{\pi^2}{6}(4C_F-C_A)^2 \, ,
\eeqa
with \cite{GV}
\beq 
c_a^{q \bar q}=C_F\left[-\frac{3}{2}+\ln\left(\frac{1-v}{v}\right)\right]
\eeq
and 
\beqa 
c_1^{q \bar q}&=&\frac{\beta_0}{4}\ln\left(\frac{\mu_R^2}{s}\right)
+\frac{1}{2}K-\frac{7}{2}C_F+\frac{\pi^2}{2}\left(C_F-\frac{C_A}{6}\right)
\nonumber \\ && 
{}+\frac{n_f}{6}\ln v- C_F\ln v \, \ln (1-v)
\nonumber \\ && \hspace{-10mm}
{}+\frac{C_F}{2T_{q \bar q}}\left[v(2+v)\ln (1-v)+(1-v)(3-v) \ln v \right.
\nonumber \\ && \hspace{10mm} \left.
{}+(3v^2+2(1-v))\ln^2 v+(2v+3(1-v)^2)\ln^2 (1-v)\right]
\nonumber \\ && \hspace{-10mm}
{}-\frac{N_c}{2 T_{q \bar q}}\left[v (1-v) \ln (1-v)
+\frac{(11-16v(1-v))}{6}\ln v \right.
\nonumber \\ && \hspace{10mm} \left.
{}+\frac{(2v+3(1-v)^2)}{2}\ln^2 (1-v)+\frac{v(2-v)}{2}\ln^2 v\right] \, .
\nonumber \\
\eeqa

Note that, at NNLO, we have derived all NNLL terms, i.e. 
all the ${\cal O}\left(\left[\ln(1-w)/(1-w)\right]_+\right)$ terms,
by matching with the exact NLO cross section in Ref. \cite{GV}. 
The scale-dependent terms at that accuracy have also been derived by 
the expansion of the resummed cross section, giving the same result.
As an additional check, an explicit variation of the hadronic cross
section with respect to scale yields zero up to $[1/(1-w)]_+$ terms.
From the expansion of the resummed cross section we can also derive 
the $[1/(1-w)]_+$ terms containing squared logarithms of the scale.
These terms in the coefficient ${\tilde c}_2^2(qg)$ are
\beqa
&& \ln^2\left(\frac{\mu_F^2}{s}\right) \, (C_F+C_A)\left[\frac{3}{4}C_F
-C_A\ln\left(\frac{1-v}{v}\right)+\frac{3\beta_0}{8}\right]
\nonumber \\ &&
{}-\frac{\beta_0}{2} \, (C_F+C_A) \,
\ln\left(\frac{\mu_R^2}{s}\right) \, \ln\left(\frac{\mu_F^2}{s}\right) \, ,
\eeqa
while the corresponding terms in the coefficient 
${\tilde c}_2^2(q {\bar q})$ are
\beq
\ln^2\left(\frac{\mu_F^2}{s}\right) \,
C_F\left[C_F\left(3-2\ln\left(\frac{1-v}{v}\right)\right)
+\frac{\beta_0}{4}\right]
-C_F \beta_0 \ln\left(\frac{\mu_R^2}{s}\right) \, 
\ln\left(\frac{\mu_F^2}{s}\right) \, .
\eeq

Finally, we note that all gauge dependence cancels explicitly in
the NNLO expansions, as expected.

\end{document}